\documentclass[a4paper,11pt]{article}
\usepackage[toc,page]{appendix}
\usepackage{graphicx}
\usepackage{colortbl}
\usepackage{amsmath}
\usepackage{subfigure}
\usepackage{mathtools}
\usepackage[utf8]{inputenc}
\usepackage{float}
\usepackage[normalem]{ulem}
\usepackage{epstopdf}
\usepackage{amsfonts,amsmath,amssymb}
\usepackage{hyperref}

\usepackage{epsfig,multicol,bbm}
\usepackage{color}
\usepackage[dvipsnames]{xcolor}

\definecolor{darkblue}{rgb}{0.0, 0.0, 0.55}
\definecolor{grey}{rgb}{0.57, 0.64, 0.69}
\definecolor{lightbrown}{rgb}{0.71, 0.4, 0.11}
\newcommand{\tcb}{\textcolor{blue}}

\newcommand{\be}{\begin{equation}}
\newcommand{\ee}{\end{equation}}

 \newcommand{\rbm}[1]{{\color{red}\bf [Robb: #1]}}

\newcommand\fverb{\setbox\pippobox=\hbox\bgroup\verb}
\newcommand\fverbit{\egroup\item[\fbox{\unhbox\pippobox}]}

\newbox\pippobox

\begin{document}

 \title{Charged Black Holes in Einsteinian Quartic Gravity}
 \author{S. N. Sajadi\thanks{Electronic address: naseh.sajadi@gmail.com}, Leila Shahkarami
 \thanks{Electronic address: l.shahkarami@du.ac.ir}, Farid Charmchi
 \thanks{Electronic address: charmchi@ipm.ir}, S. H. Hendi\thanks{Electronic address: hendi@shirazu.ac.ir}
\\
\small Department of Physics, School of Science, Shiraz University, Shiraz 71454, Iran\\
\small Biruni Observatory, School of Science, Shiraz University, Shiraz 71454, Iran\\
\small School of Physics, Damghan University, Damghan, 41167-36716, Iran\\
\small School of Particles and Accelerators, Institute for Research in Fundamental Sciences (IPM),\\ P.O.Box 19395-5746, Tehran, Iran}

\maketitle
\begin{abstract}
In this paper, we studied Einsteinian quartic gravity minimally coupled to electrodynamics in four dimensions. First, by variation action, we obtain the field equations, and by integration, we obtain a nonlinear third-order differential equation for the metric function and as well as the electromagnetic potential. Then, in the context of the Maxwell field, we discussed
the conditions under which the charged black hole exists. Then, we have demonstrated the thermodynamics and stability of the solution for the case of positive coupling of quartic theory. Finally, we showed that the charged black hole solutions of EQG (unlike GR and like ECG) have no inner horizon and do not conform to the extremal bound of GR. Also, the uniqueness of BH solutions in this theory does not work anymore.
\end{abstract}

\section{Introduction}
Einstein's general relativity is one of the most successful theories of physics and its predictions have been well established in recent years \cite{EventHorizonTelescope:2019dse}-\cite{LIGOScientific:2020iuh}. Despite these successes, it has had problems on various scales. In cosmology, GR is faced with the problems of dark matter and dark energy and does not unify with quantum on a small scale. As a result, the theory needs to be modified. One of the most natural modifications is to add higher curvature terms to Einstein-Hilbert's action \cite{Sotiriou:2008rp}-\cite{Stelle:1976gc}. But, most higher curvature theories have ghost degrees of freedom in their propagator. The Lovelock class of theories is ghost-free but in four dimensions are trivial. In recent years, a new class of higher derivative theories has been discovered that is ghost-free and in four dimensions neither topological nor trivial known as Generalized Quasi-Topological Gravity \cite{Bueno:2016xff}, \cite{Hennigar:2017ego}, \cite{Ahmed:2017jod}. An important feature of these theories is that their field equations admit static, spherically symmetric solutions with a single metric function. The first such theory is cubic \cite{Bueno:2016xff} and the second one is quartic in curvature \cite{Ahmed:2017jod}. The Einstein quartic gravity has six quartic curvature combination Lagrangians which lead to the introduction of six different coupling constants. However, the condition of spherical symmetry leads to a
degeneracy between coupling constants.
 In EQG, the second-order nonlinear differential equations for the single metric function are obtained by integration of the field equations obtained from the variation of the action. This field equation analytically can not be solved but approximately using continued fraction expansions in \cite{Khodabakhshi:2020hny} has been solved and the black hole and black brane solutions have been studied. Black holes correspond to extreme deformations of spacetime geometry and can exist even without matter. These objects are the simplest objects in the world because using the uniqueness theorem they are uniquely determined by their mass, charge, and angular momentum. In \cite{Ahmed:2017jod}, the black brane solutions in arbitrary dimensions and, the thermodynamics of four-dimensional asymptotically flat black hole solutions were studied and they found the first law is satisfied. In the paper \cite{Khodabakhshi:2020hny}, the basic tests of Einstein Quartic Gravity using the continued fraction expansion for black hole solution, including the solar system tests, the motion of particles around a black hole, and the properties of a black hole shadow have been studied. They found the theory is compatible with solar system tests in the large values of the coupling constants. 
In this paper, following the paper \cite{Frassino:2020zuv}, we analytically obtain the thermodynamical quantities using a local analysis around the horizon of black holes, thermodynamical stability, nonuniqueness of charged black holes, and the existence
of horizonless solutions.

The paper is organized as follows: In the next section, we first
review Einsteinian quartic gravity and we get the field equations of the theory. Next, we obtained the conditions in which the black hole could exist and their thermodynamical stability.
Then, we obtained the singular solutions for pure Einsteinian quartic gravity. We finish the paper with some concluding remarks.

\section{Basic Equations}\label{sec2}
\begin{figure}[H]\hspace{0.4cm}
\centering
\subfigure[]{\includegraphics[width=0.3\columnwidth]{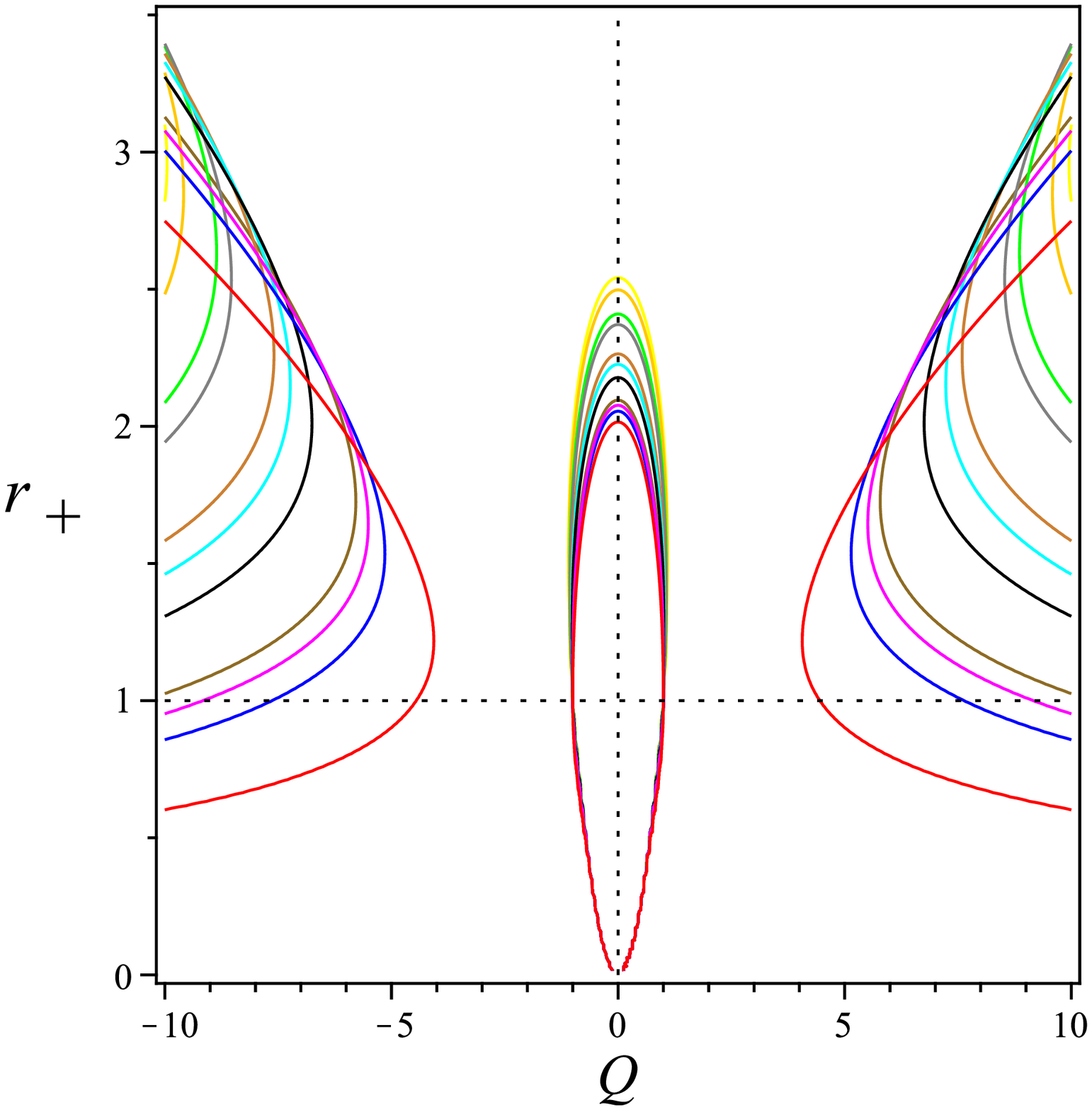}}
\subfigure[]{\includegraphics[width=0.3\columnwidth]{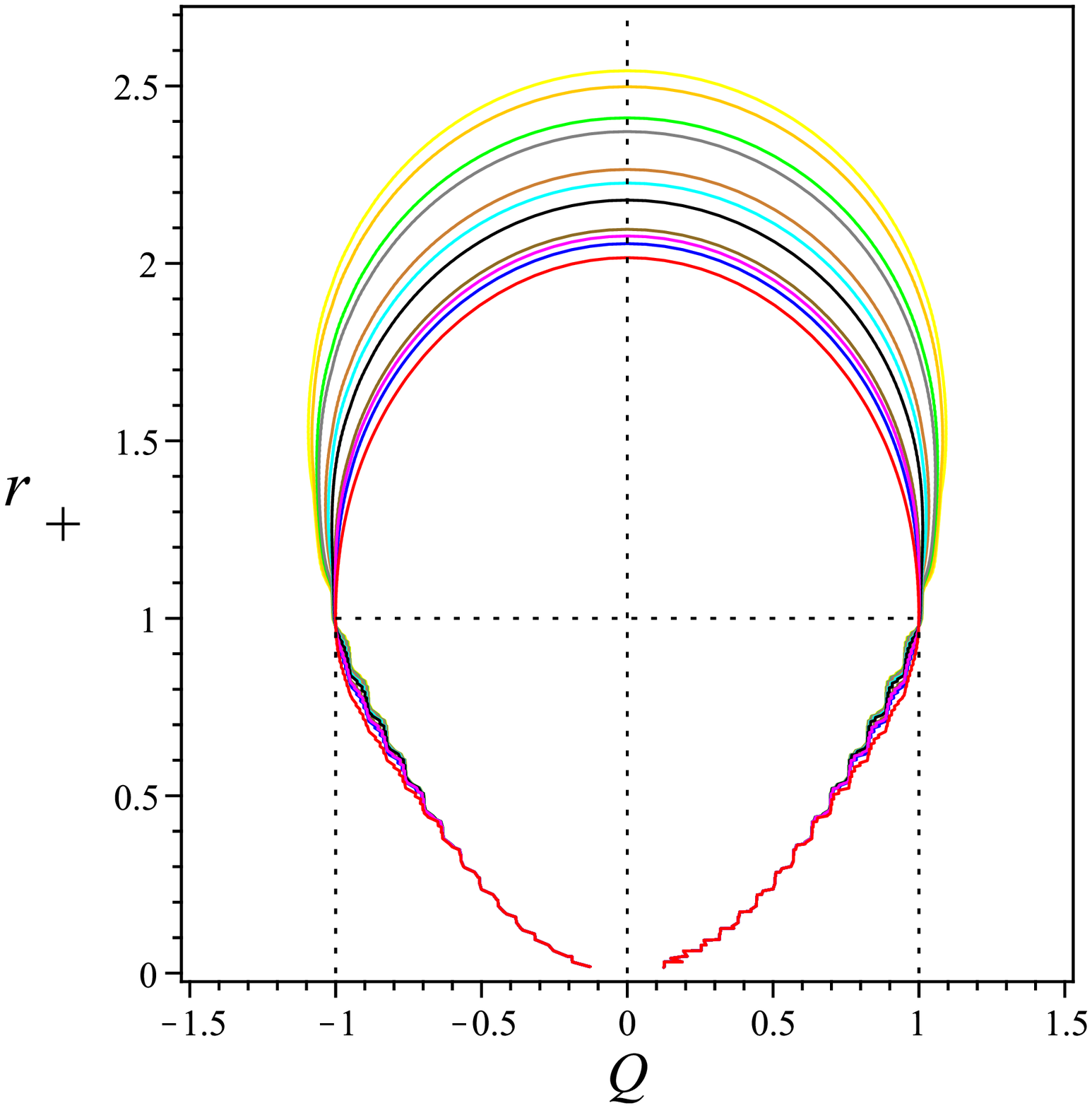}}
\subfigure[]{\includegraphics[width=0.3\columnwidth]{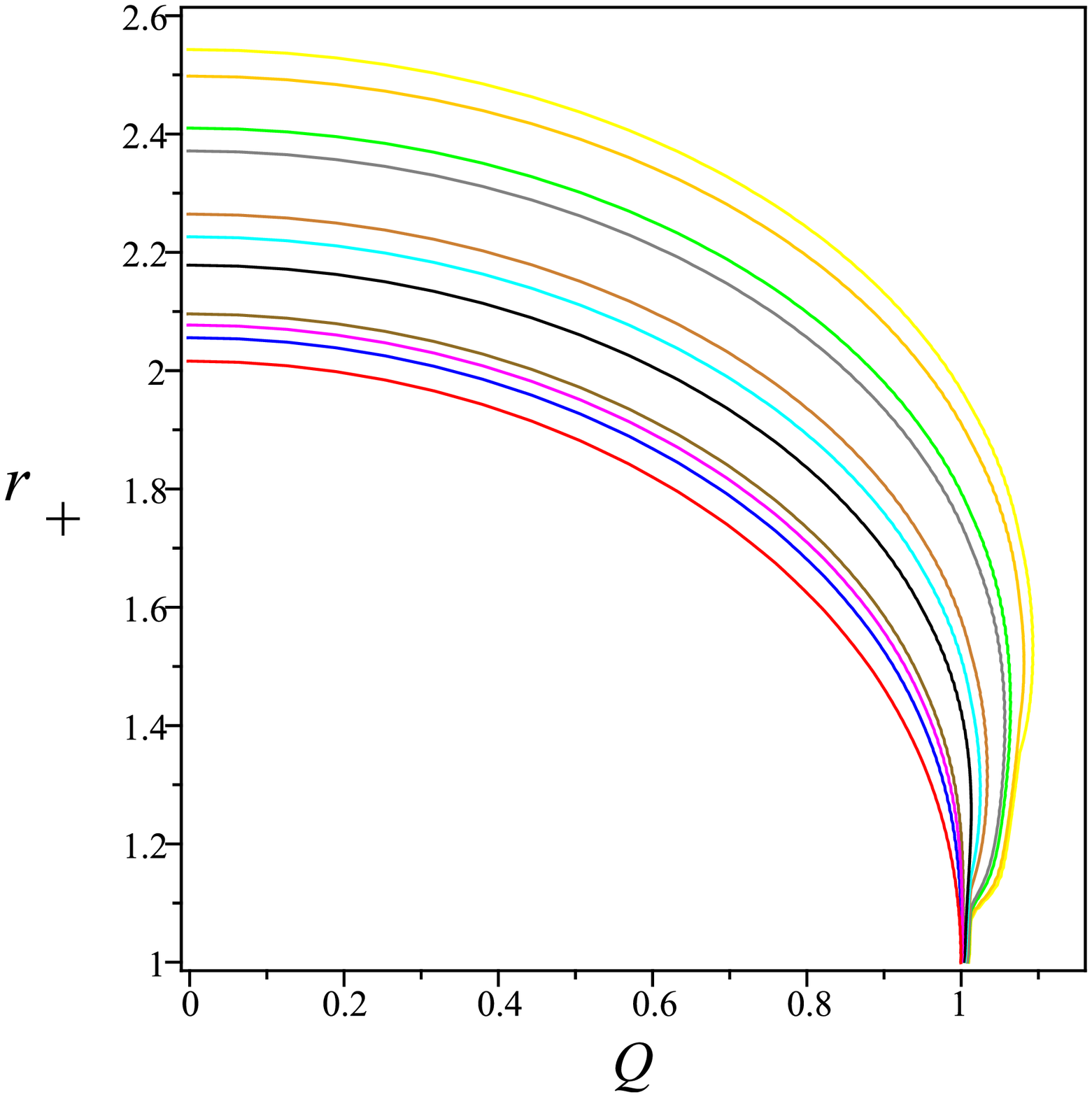}}
\caption{The plots of horizon radius $r_{+}$ in terms of $Q$ for large $\mathcal{K}=\textcolor{red}{0}, 0.1, 0.5, 1, 10, 20, \textcolor{orange}{100}$ (left). Zoom in around the middle part of the left figure (middle). Zoom in around the right part of the middle figure (right). (The quantities are in units of the mass $M$)}
\label{rrqqplot00}
\end{figure}

\begin{figure}[H]\hspace{0.4cm}
\centering
\subfigure[]{\includegraphics[width=0.3\columnwidth]{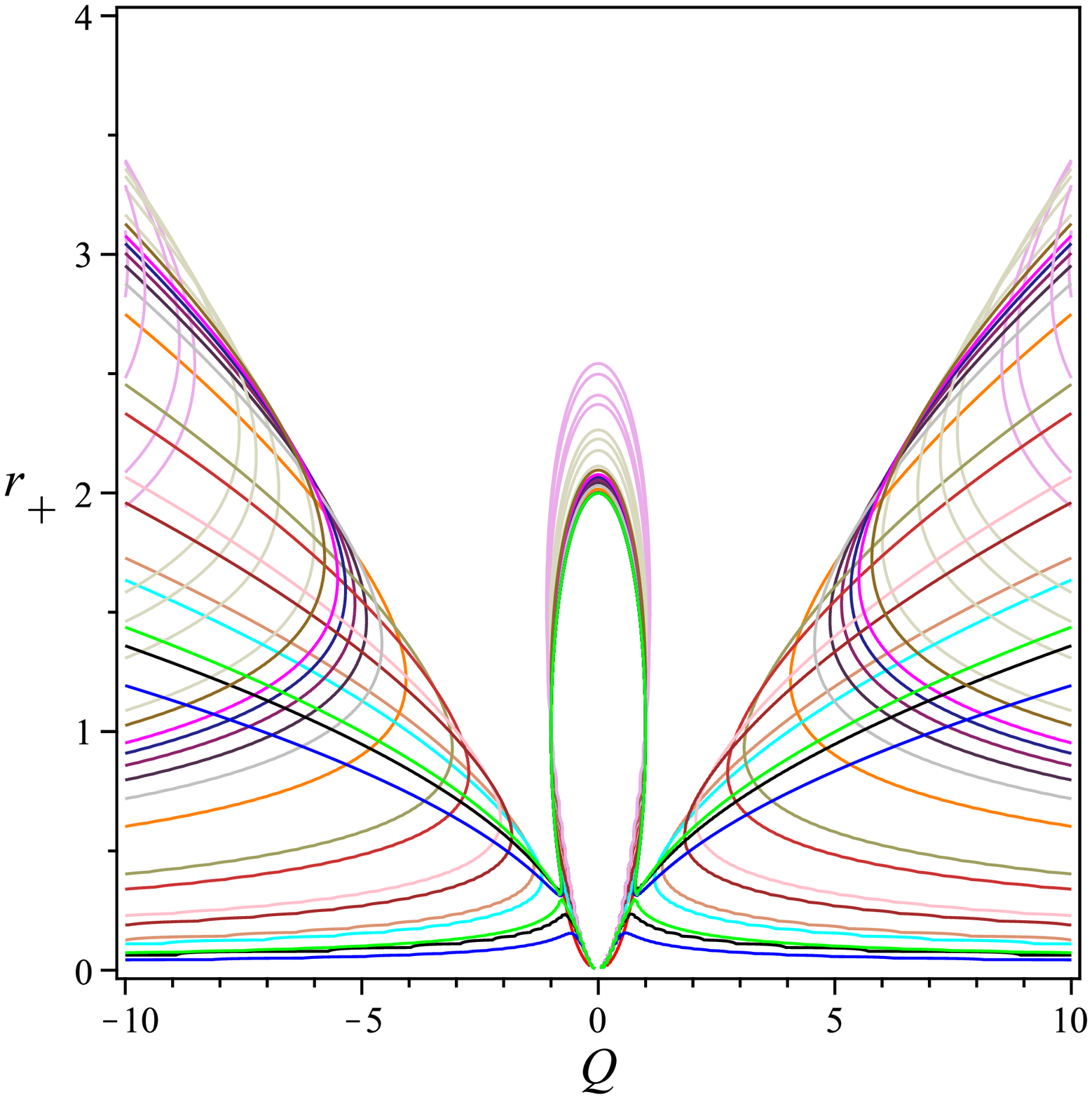}}
\subfigure[]{\includegraphics[width=0.3\columnwidth]{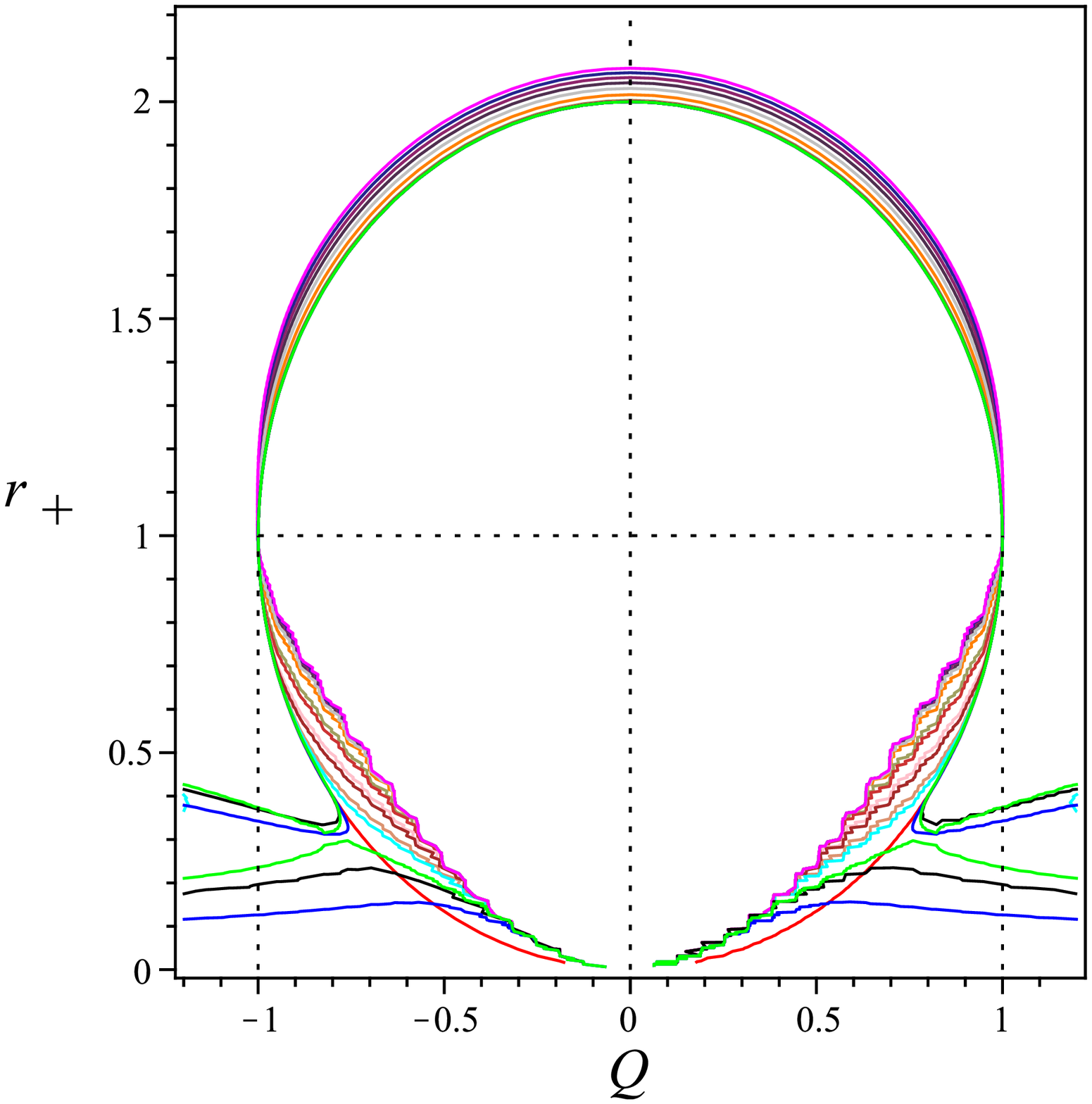}}
\subfigure[]{\includegraphics[width=0.3\columnwidth]{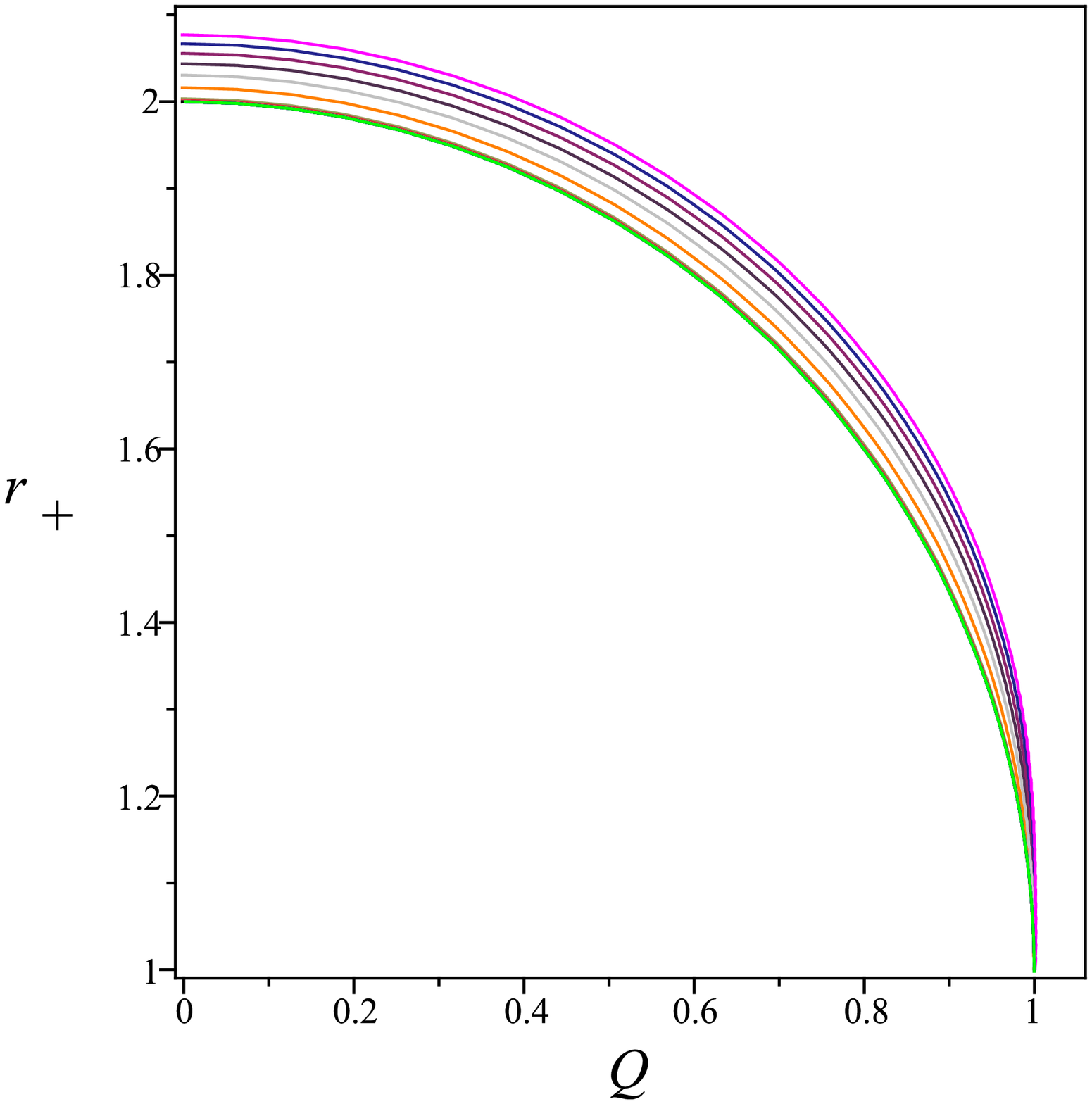}}
\caption{The plots of horizon radius $r_{+}$ in terms of $Q$ for small $\mathcal{K}=\textcolor{red}{0}, 0.00001, 0.00005, 0.0001, 0.0005, 0.001, 0.005, 0.01, 0.05, 0.1, 0.5, \textcolor{cyan}{1}$ (left). Zoom in around the middle part of the left figure (middle). Zoom in around the right part of the middle figure (right). (The quantities are in
units of the mass $M$)}
\label{rrqqplot11}
\end{figure}

The action of Einstein quartic gravity (EQG) coupled to nonlinear electrodynamics in 4D can be written as
\begin{equation}
S=\int d^{4}x\sqrt{-g}\mathcal{L},
\end{equation}
with Lagrangian
\begin{align}\label{eq1}
\mathcal{L}&=R-2\Lambda-\sum \hat{\alpha}_{i}\hat{L}^{i}+\dfrac{1}{4}\left(-F\right)^{n},
\end{align}
where $F=F_{a b}F^{a b}$ and $F_{a b}=2A_{[a;b]}$ is the electromagnetic tensor, $\hat{L}_{i}$ are quasi-topological Lagrangian densities, whose analytical expressions are given in \cite{Ahmed:2017jod}, $\Lambda$ is cosmological constant and
 $\hat{\alpha}_{i}$ are coupling constants of the theory. 
Using the variational principle, one can find the following equation of motion
\begin{align}\label{eqqfield}
E_{a b}&=P_{a c d e}R_{b}{}^{c d e}-\dfrac{1}{2}g_{ab}\mathcal{L}-2\nabla^{c}\nabla^{d}P_{a c d b}-2T_{a b}=0,\\
&\nabla_{a}\left(F^{n-1}F^{a b}\right)=0,\;\;\;\; P_{a b c d}=\dfrac{\partial \mathcal{L}}{\partial R^{a b c d}}.
\end{align}
Here, we consider the following spherically symmetric and static
line element for describing the geometry of spacetime
\begin{align}\label{metform}
ds^{2}&=-N^{2}(r)f(r)dt^{2}+\dfrac{dr^{2}}{f(r)}+r^2\left(d\theta^{2}+\sin^{2}(\theta)d\phi^{2}\right),\\
A(r)&=A_{0}(r)dt.
\end{align}
Varying the action (\ref{eq1}) with respect to $N(r)$, $f (r)$ and $A_{0}(r)$, we are left with three field equations.
One of these field equations which arises from variation with respect to $f (r)$, is satisfied by $N(r) = c$. Then, we have two field equations as follows
\begin{align}
&5r^{6}(2n-1)A_{0}^{\prime}{}^{2n}+20r^{5}f^{\prime}+20r^4(\Lambda r^2+f-1)-\mathcal{K}[24r^2f^{\prime}ff^{\prime\prime\prime}(2-2f+rf^{\prime})+48r^{2}ff^{\prime\prime 2}(1-\nonumber\\
&f+f^{\prime}r)+16rf^{\prime 3}(7f-1)+144ff^{\prime 2}(1-f)-48rff^{\prime}f^{\prime\prime}(3rf^{\prime}-4f+4)-2r^2f^{\prime 4}]=0,\label{fieldequation}\\
&r\left(2n-{1}\right)A_{0}^{\prime\prime}A_{0}^{\prime}{}^{2n-2}+2A_{0}^{\prime} {}^{2n-1}=0,\label{eqelectro}
\end{align}
where prime is the derivative to $r$ and $\mathcal{K}$ is a function of coupling constans $\hat{\alpha}_{i}$.
 The solution of electrodynamic field equation (\ref{eqelectro}) is given as
\begin{equation}\label{eqa0}
A_{0}(r)=c_{2}+\dfrac{(2n-1)r\left(\dfrac{c_{1}}{r}\right)^{\frac{2}{2n-1}}}{-3+2n}.
\end{equation}
Inserting \eqref{eqa0} into the modified Einstein field equations \eqref{fieldequation} and by integrating it, one can get a nonlinear differential equation for $f(r)$ as
\begin{align}\label{eqfiled2}
&72(2n-3)\left(-\dfrac{r^2}{4}f^{\prime 4}-\dfrac{r}{3}(2+f)f^{\prime 3}+2f(f-1)f^{\prime 2}+rff^{\prime}f^{\prime\prime}(2-2f+rf^{\prime})\right)\mathcal{K}-\nonumber\\
&60\left(n-\dfrac{1}{2}\right)^{2}c_{1}^{\frac{4n}{2n-1}}r^{\frac{8n-6}{2n-1}}-60(2n-3)\left(\dfrac{\Lambda r^3}{3}+fr-r+2M\right)r^3=0,
\end{align}
the parameter $M$ is the integration constant which is related to the mass of the spacetime.
The equation \eqref{eqfiled2}, is not solved analytically. Therefore, we solve it approximately. In the approximate calculation method, one first gets the solution close to the horizon of the black hole.
So, expanding the function $f(r)$ around the event horizon $ r_{+} $
\begin{align}\label{eq7}
f(r)  &= f_{1}(r-r_{+})+f_{2}(r-r_{+})^{2}+f_{3}(r-r_{+})^{3}+...
\end{align}
and then inserting these expressions into equations (\ref{eqfiled2}), one can consider the two lowest order equations as
\begin{align}\label{eq9}
&15(2n-1)^{2}c_{1}^{\dfrac{4n}{2n-1}}r_{+}^{\dfrac{8n-6}{2n-1}}+18(2n-3)r_{+}\left(\mathcal{K}r_{+}f_{1}^{4}+\dfrac{8}{3}\mathcal{K}f_{1}^{3}+\dfrac{10}{9}\Lambda r_{+}^5+\dfrac{20}{3}Mr_{+}^2-\dfrac{10}{3}r_{+}^{3}\right)=0,\\
&-5(2n-1)c_{1}^{\dfrac{4n}{2n-1}}r_{+}^{\dfrac{8n-6}{2n-1}}+20r_{+}^{4}-16\mathcal{K}f_{1}^{3}r_{+}-20f_{1}r_{+}^{5}-2\mathcal{K}f_{1}^{4}r_{+}^{2}-20\Lambda r_{+}^{6}=0,
\end{align}
by substituting $c_{1}=\sqrt{2Q}$ in the above equations, and solving them simultaneously the following solutions for $Q$ and $r_{+}$ can be reached:
\begin{align}
Q^{2}&=\dfrac{(2n-3)^{2}}{240n^{4}(2n-1)^{2}\Lambda^{2}}\left(\dfrac{20\mathcal{A}^{6}}{\mathcal{B}}\right)^{\dfrac{1}{n}}\times\nonumber\\
&\dfrac{65536\mathcal{A}\mathcal{J}n^{6}+65536\mathcal{G}n^{5}
+65536\mathcal{H}n^{4}-172800f_{1}\mathcal{N}n^{3}-
1166400f_{1}^{3}\mathcal{P}n^{2}+340200\mathcal{S}f_{1}^{5}n+
18225f_{1}^{7}\mathcal{X}}{128\mathcal{A}\mathcal{T}n^{4}+
128\mathcal{U}n^{3}+128\mathcal{V}n^{n}+1080f_{1}\mathcal{W}n+135f_{1}^{3}n},\label{eqq155}\\
r_{+}&={\it RootOf}( 20n{\it \_Z}^{5}\Lambda+( 30nf_{1}-
15f_{1}) {\it \_Z}^{4}-30{\it \_Z}^{3}+(-60nM+90M) {\it \_Z}^{2}+(12\mathcal{K}f_{1}^{4}-\nonumber\\
&6n\mathcal{K}f_{1
}^{4}) {\it \_Z}+24\mathcal{K}f_{1}^{3}),\label{eqqrplus}
\end{align}
where $\mathcal{A}$, $\mathcal{B}$, $\mathcal{G}$... are provided in appendix \ref{appp1}, and ${\it RootOf}$ is a command used as a placeholder for roots of equations in maple \cite{maple}. The other constants in \eqref{eq7}, provided in appendix \ref{appp2}.
In the follwoing first we consider the case $n=1$ and $\mathcal{K}>0$. Therefore, by setting $M=1,\Lambda=0$ from (\ref{eq9})-\eqref{eqqrplus} we have
\begin{align}
&-\dfrac{3\mathcal{K}f_{1}^{4}}{10r_{+}}-\dfrac{4\mathcal{K}f_{1}^{3}}{5r_{+}^{3}}+\dfrac{Q^{2}}{r_{+}}-2+r_{+}=0,\label{eqqfi1}\\
&-\dfrac{\mathcal{K}f_{1}^{4}}{10r_{+}^2}-\dfrac{4\mathcal{K}f_{1}^{3}}{5r_{+}^{3}}-\dfrac{Q^2}{r_{+}^{2}}+1-r_{+}f_{1}=0,\label{eqqfi2}
\end{align}
and solving them for $Q$ and $r_{+}$ one can get
\begin{align}
Q^{2}&=\dfrac{3\mathcal{K}\mathcal{E}f_{1}^{4}+8\mathcal{K}f_{1}^3-10\mathcal{E}^{3}+20\mathcal{E}^2}{10\mathcal{E}},\\
r_{+}&={\it RootOf}( 2{f_{1}}^{4}\mathcal{K}{\_Z}+8{f_{1}}^{3}\mathcal{K}+10{{ \_Z}}^{
2}-10{{\_Z}}^{3}+5{{\_Z}}^{4}f_{1}),
\end{align}
here $\mathcal{E}$ is
\begin{equation}
\mathcal{E}={\it RootOf}\left( 2\mathcal{K}{f_{1}}^{4}{\_Z}+8\mathcal{K}{f_{1}}^{3}-10{{\_Z}}^{3}+10{{\_Z}}^{2}+5{{\_Z}}^{4}f_{1} \right).
\end{equation}
In figures \eqref{rrqqplot00} and \eqref{rrqqplot11} the behavior of horizon radius $r_{+}$ in terms of $Q$ for $f_{1}>0$ and for different positive large and small values of $\mathcal{K}$ are illustrated, respectively. The innermost red curve corresponds to $\mathcal{K}=0$ (RN-solution), and the black dot at the end of the curve indicates the extremal solution, for which $r_{+}=Q=M=1$.
The figures show that there are the real solutions in which $Q>1$, suggesting that charged BHs in EQG need not comply with the extremality condition of Einstein-Maxwell theory. One can find a finite interval $ 1<Q<Q_{max}$ where there are two real solutions with $f_{1}>0$. 
$Q_{max}$ is a function of $\mathcal{K}$, which can be easily determined by finding
the root of $dQ/dr_{+}$. So, for $\mathcal{K}>\mathcal{K}_{b}=0.1538716564$ there are the real solutions over extremal regime with positive temperature ($T=4\pi f_{1}$). \\

Solving the equations \eqref{eqqfi1} and \eqref{eqqfi2} for $T$ and $M$, one can get
\begin{align}\label{eqtemp}
&T=\dfrac{1}{4\pi}\left[{\it RootOf} \left( 5Q{r_{+}}+512{\pi }^{4}{{\_Z}}^{4}\mathcal{K}{ 
r_{+}}+1024{\pi }^{3}{{\_Z}}^{3}\mathcal{K}-20{{r_{+}}}^{3}+80{{r_{+}}}
^{4}\pi {\_Z} \right)\right],\\
&M=-\dfrac{1}{5r_{+}^{2}}[2048\pi^{3}({\it RootOf}(5Qr_{+}
+512\pi^{4}{\_Z}^{4}\mathcal{K}{r_{+}}+1024\pi^{3}{\_Z}^{
3}\mathcal{K}-20r_{+}^{3}+80r_{+}^{4}\pi{\_Z}) 
) ^{3}\mathcal{K}+\nonumber\\
&5Qr_{+}-40r_{+}^{3}+120\pi r_{+}^{4}
{\it RootOf}(5Q{r_{+}}+512{\pi }^{4}{\_Z}^{4}\mathcal{K}{
r_{+}}+1024\pi^{3}{\_Z}^{3}\mathcal{K}-20r_{+}^{3}+80
r_{+}^{4}\pi {\_Z})],
\end{align}
and by using Wald's formula, it is possible to evaluate the entropy of these black hole solutions \cite{Wald1}, \cite{Wald2} 
\begin{equation}
S=-2\pi\int d^{2}x \sqrt{h}\dfrac{\delta \mathcal{L}}{\delta R_{a b c d}}\epsilon_{a b}\epsilon_{c d}=\pi r_{+}^{2}\left[1+\dfrac{32\pi^2 \mathcal{K} T^2}{3r_{+}^{4}}\left(3+4\pi r_{+}T\right)\right],
\end{equation}
where $h$ is the determinant of the induced metric on the horizon and $\epsilon_{a b}\epsilon^{a b}=-2$.
The above relation reduces to $\pi r_{+}^{2}$ if the
higher-order terms disappear ($\mathcal{K}=0$).
Now, we turn to calculate the electric potential. The electric potential $\Phi$, measured at infinity with respect to horizon is defined by
\begin{equation}
\Phi=\left. A_{\mu}\xi^{\mu}{}\right\vert_{r\to \infty}-\left. A_{\mu}\xi^{\mu}{}\right\vert_{r\to r_{+}}.
\end{equation}
Using $\xi=\partial_{t}$ and (\ref{eqa0}), one can find
\begin{equation}\label{elecphi}
\Phi=\dfrac{(2n-1)r\left(\dfrac{c_{1}}{r}\right)^{\frac{2}{2n-1}}}{3-2n}.
\end{equation}
By using the thermodynamical quantities \eqref{eqtemp}-\eqref{elecphi}, it is easy to check that the first law of thermodynamics ($dM=TdS+\Phi dQ$) is satisfied similar to what happens in quadratic and cubic gravity \cite{Sajadi:2020axg}-\cite{Hennigar:2018hza}.

\begin{figure}[H]\hspace{0.4cm}
\centering
\subfigure[]{\includegraphics[width=0.45\columnwidth]{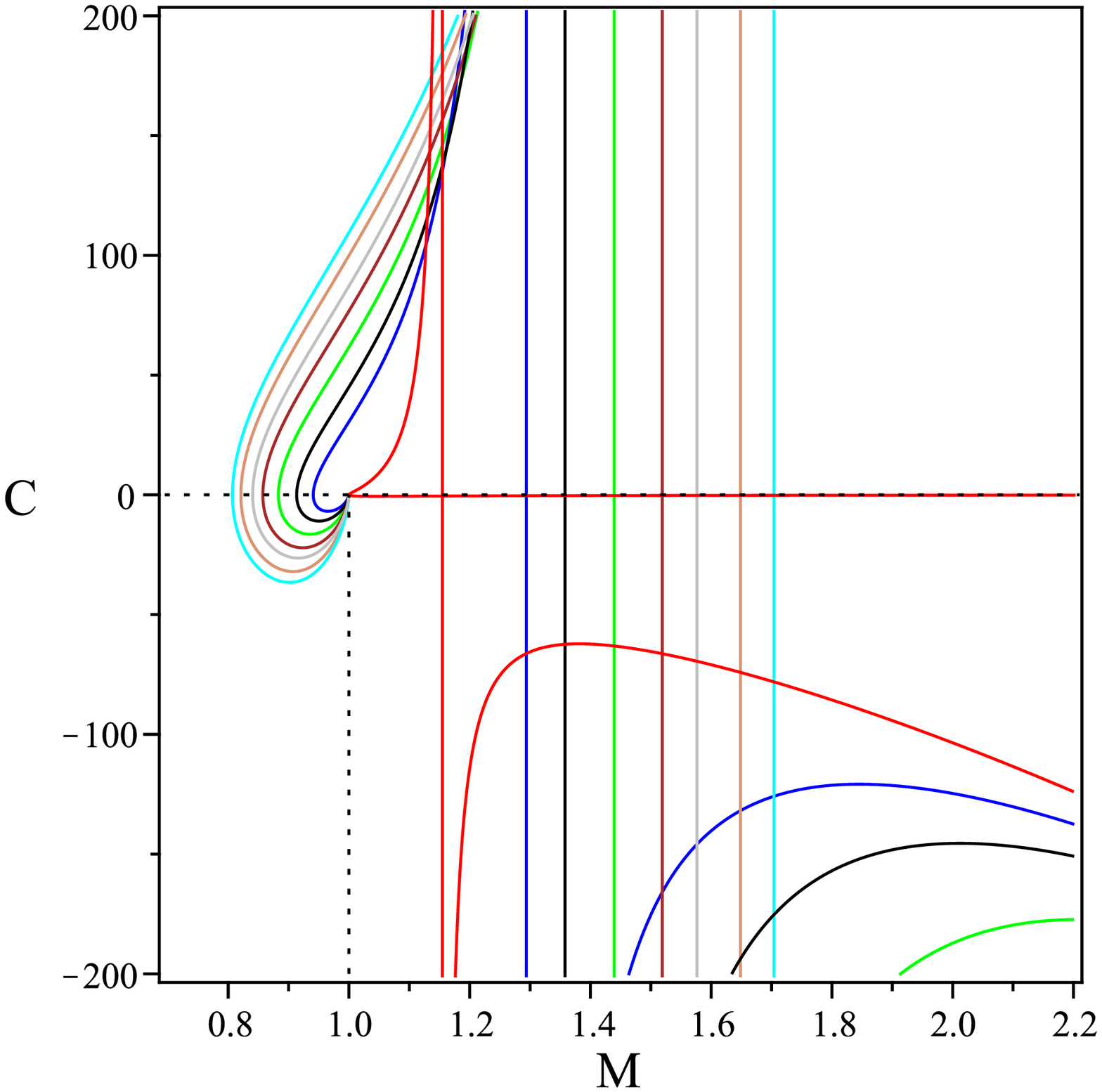}}
\subfigure[]{\includegraphics[width=0.45\columnwidth]{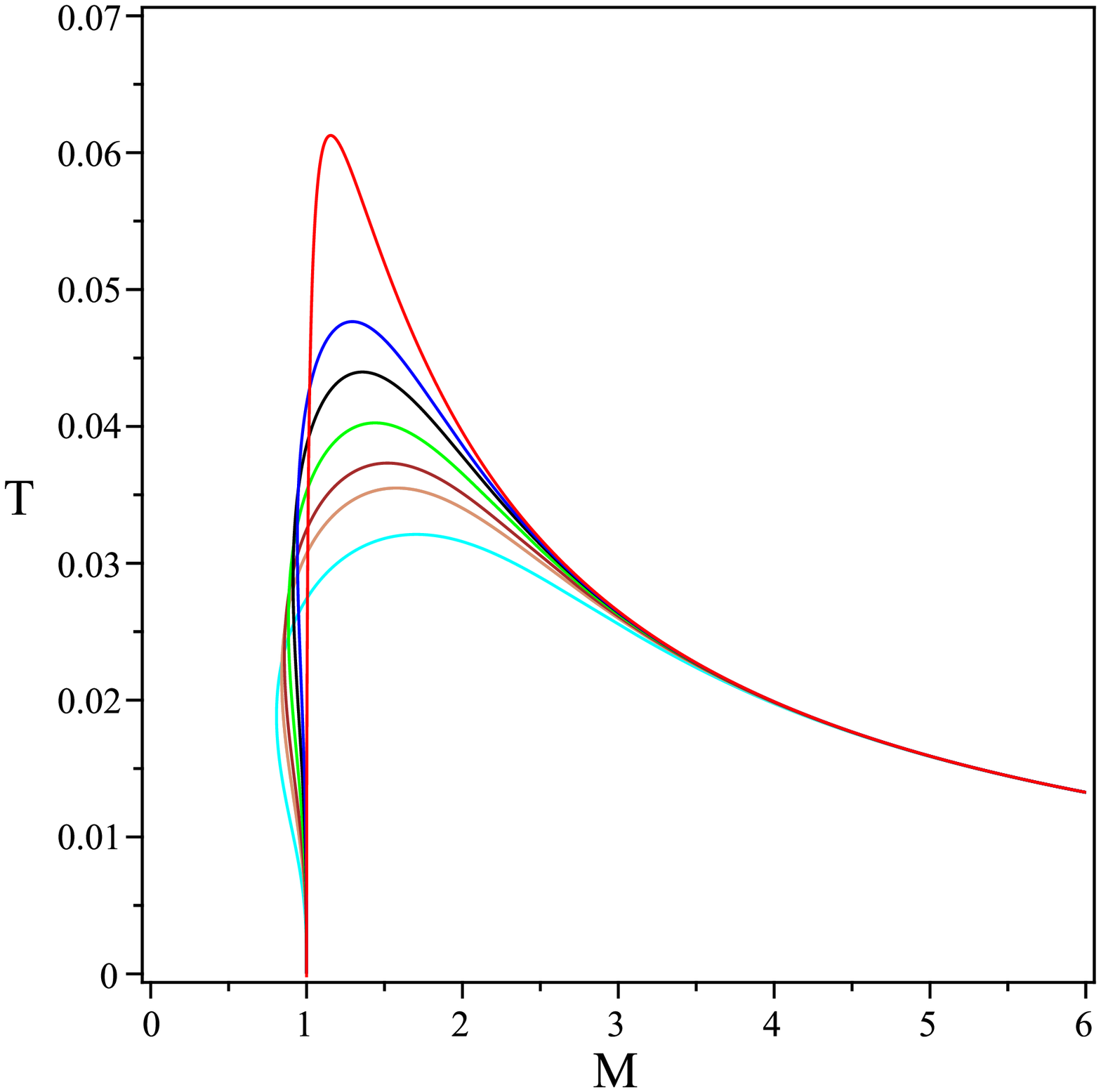}}
\caption{The plots of $C$ in terms of $M$ for $\mathcal{K}=\textcolor{red}{0}, 1, 2, 4, 7, 10,15, 40, 60, 80, \textcolor{green}{100}$ (left). The plots of $T$ in terms of $M$ for $\mathcal{K}=\textcolor{red}{0}, 1, 2, 4, 7, 10,15,\textcolor{pink}{20}$ (right). (The quantities are in
units of the charge $Q$)}
\label{rqp}
\end{figure}

\begin{figure}[H]\hspace{0.4cm}
\centering
\subfigure{\includegraphics[width=0.5\columnwidth]{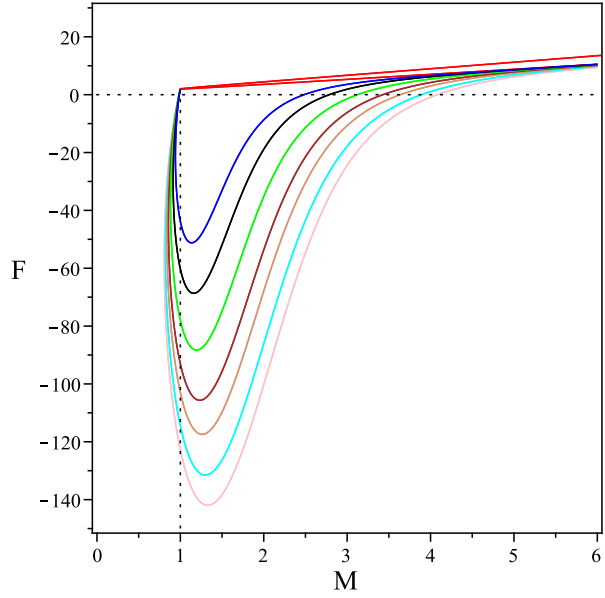}}
\caption{The plots of $F$ in terms of $M$ for $\mathcal{K}=\textcolor{red}{0}, 1, 2, 4, 7, 10,15,\textcolor{pink}{20}$.}
\label{ffrrplot0}
\end{figure}

Now, we would like to check the thermodynamical stability of the four-dimensional asymptotically flat black hole solutions of EQG. Useful information regarding the local stability can be extracted
from the behavior of the specific heat at a constant charge, $C_{Q} = T(\partial S/\partial T)_{Q}$ (canonical ensemble).
It can be seen from the figure (\ref{rqp}a) for $M<1$, that there are two BH branches
with opposite signs specific heats. The larger black holes display positive specific heat, while smaller black holes have negative specific heat. This shows that the larger BH is locally stable. In figure (\ref{rqp}b), the corresponding temperatures as a function of $M$ (at a fixed charge) are shown. This figure shows that the temperature is positive for two branches of solutions and as the mass parameter of quartic terms 
grows, the temperature and $T_{max}$ of black holes decreases.
In the figure (\ref{ffrrplot0}), in order to look at the global stability, the behavior of free energy ($F=M-TS$) has been shown. As can be seen, the free energy of larger BHs is lower which shows that globally is stable.

\begin{figure}[H]\hspace{0.4cm}
\centering
\subfigure[]{\includegraphics[width=0.45\columnwidth]{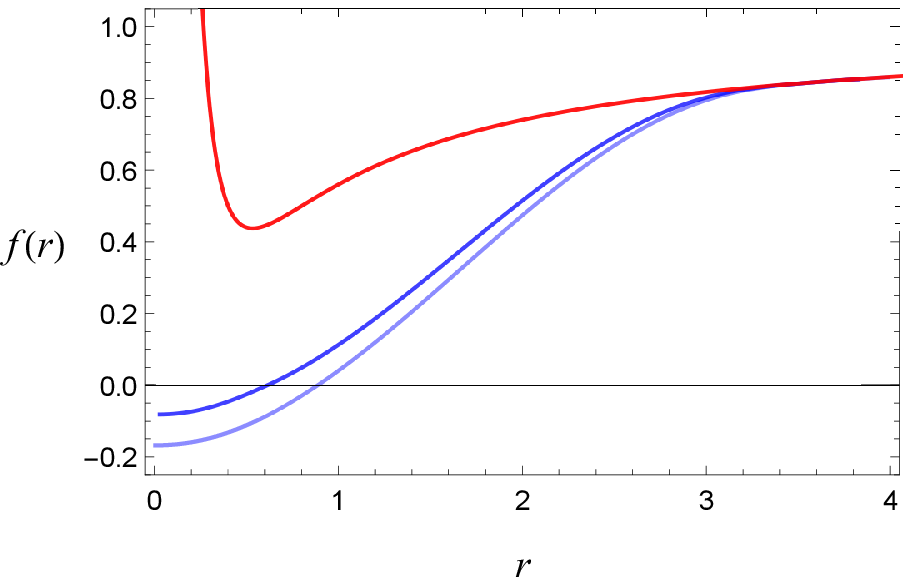}}
\subfigure[]{\includegraphics[width=0.45\columnwidth]{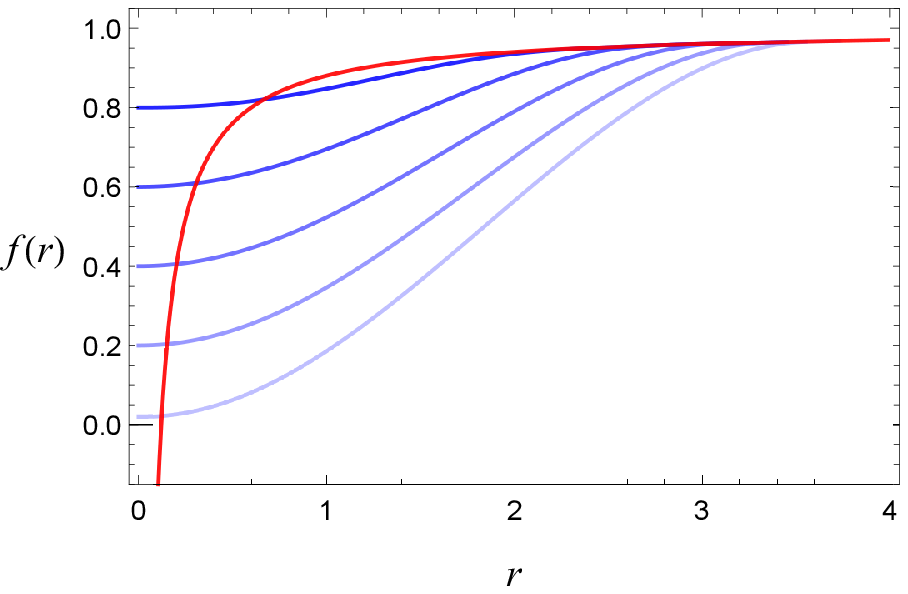}}
\caption{The plots of the $f(r)$ for two distinct BH solutions with tha same $M$ and $Q=1.3$ and $\mathcal{K}=10$ (left). The plots of the $f(r)$ for naked singularities for the same $M$ and different $c_{0}=0.8,0.6,0.4,0.2,0.02$ (right). The red line is the corresponding RN over-extremal solution for $Q=1.3, \mathcal{K}=0$. (The quantities are in units of the mass $M$)}
\label{ffrrplot00}
\end{figure}

As we have seen for $\mathcal{K}>\mathcal{K}_{b}$, there are the black hole solutions with the same $M$ and $Q$ but different event horizon radii. Here, we show that both solutions for $r_{+}$ are associated with asymptotically flat spacetimes. So, we consider the large $r$ limit of metric as 
\begin{align}\label{eqqqhfo}
 f(r)=1-\dfrac{2M}{r}+\epsilon F(r)
\end{align}
by inserting the expansions into the field equations  (\ref{eqfiled2}), one can get 
\begin{align}
F^{\prime\prime}-\dfrac{(5r-8M)}{r(r-2M)}F^{\prime}-\dfrac{(688\mathcal{K}M^{3}-144M^{2}\mathcal{K}r+5r^{9})}{72M^{2}\mathcal{K}r^2(r-2M)}F
=\dfrac{(3104M^{4}
\mathcal{K}-1728M^{3}\mathcal{K}r+5Q^{2}r^{8})}{288\mathcal{K}M^{2}r^{3}(2M-r)},
\end{align}
the solution for homogenous equation at large $r$ reads
\begin{equation}
F_{h}(r)=Ar^{3}I_{\frac{3}{4}}\left(\dfrac{\sqrt{5}r^4}{48M\sqrt{\mathcal{K}}}\right)+Br^{3}K_{\frac{3}{4}}\left(\dfrac{\sqrt{5}r^4}{48M\sqrt{\mathcal{K}}}\right)
\end{equation}
$I_{\nu}(x)$ and $K_{\nu}(x)$ are the first and second kinds of modified Bessel functions. To obtain the special solution we assume
\begin{equation}
F_{s}(r)=\sum_{n=2}\dfrac{F_{n}}{r^{n}}=\dfrac{F_{2}}{r^2}+\dfrac{F_{3}}{r^3}+...
\end{equation}
solving order by order for $F_{n}$, finally one can get
{\begin{align}\label{eqfasympt}
&f(r)=1-\dfrac{2M}{r}+\dfrac{Q^2}{r^2}-\dfrac{432\mathcal{K}M^{3}}{5r^9}+\dfrac{4\mathcal{K}M^2(194M^2+324Q^2)}{5r^{10}}-\dfrac{104\mathcal{K}MQ^{2}(26M^2+12Q^{2})}{5r^{11}}+\nonumber\\
&\dfrac{96\mathcal{K}Q^{4}(4Q^2+35M^2)}{5r^{12}}-\dfrac{432\mathcal{K}MQ^{6}}{r^{13}}+Ar^{3}I_{\frac{3}{4}}\left(\dfrac{\sqrt{5}r^4}{48M\sqrt{\mathcal{K}}}\right)+Br^{3}K_{\frac{3}{4}}\left(\dfrac{\sqrt{5}r^4}{48M\sqrt{\mathcal{K}}}\right)+\mathcal{O}\left(\dfrac{1}{r^{14}}\right).
\end{align}}
Asymptotic flatness demands that we should set $A = 0$.
Using the continued fraction expansion, the near horizon expansion \eqref{eq7} can be connected to the asymptotic expansion \eqref{eqfasympt} to obtain a general metric that works perfectly for the outside of the black hole. Here,
we solve the differential equation (\ref{eqfiled2}), using the NDSolve technique in Mathematica which includes the Explicit Runge-Kutta method to obtain the solutions which asymptotically approach \eqref{eqfasympt}. In figure \eqref{ffrrplot00}, for the non-extremal horizon, $T\neq 0$ we have obtained the behavior of metric function for the same conserved charges. The figure reveals the absence of an inner horizon in both black hole spacetimes.
So the causal structure of these BHs is different from that of the RN solutions and instead is similar to that of the Schwarzschild solution. This shows that the uniqueness of black holes in EQG-Maxwell theory no longer works.

Here, we are going to focusing on the center of black hole solutions ($r=0$) and obtaining the horizonless solutions of theory. To do so, we assume
an expansion $f(r) =\sum_{i} c_{i}r^{i}$ where the coefficients $c_{i}$ can be determined by solving (\ref{eqfiled2}) order by order in powers of $r$.
So, near the origin for the case of $\Lambda=Q=0$ we have
\begin{align}\label{originmetric}
f(r)&=c_{0}+c_{2}r^2-\dfrac{5Mr^3}{36\mathcal{K}c_{0}c_{2}(c_{0}-1)}+\dfrac{(384\mathcal{K}^2c_{0}c_{2}^{5}(c_{0}-1)^2-60\mathcal{K}c_{0}c_{2}^{2}(c_{0}-1)^2-25M^2)r^4}{2304\mathcal{K}^{2}c_{0}^{2}c_{2}^{3}(c_{0}-1)^{2}}\nonumber\\
&-\dfrac{5M\left(\mathcal{K}c_{0}^{3}c_{2}^{2}-2\mathcal{K}c_{0}^{2}c_{2}^{2}+\mathcal{K}c_{0}c_{2}^{2}+\dfrac{5}{12}M^2\right)r^5}{1152\mathcal{K}^3c_{0}^{3}c_{2}^{5}(c_{0}-1)^{3}}+\mathcal{O}\left(r^6\right),
\end{align}
there is one additional free parameter compared to the expansion around a horizon and asymptotic ($c_{2}$). This parameter should be determined to get asymptotically flat solutions.
The Kretschmann scalar near the origin behaves as 
\begin{equation}
K=R_{a b c d}R^{a b c d}\approx \dfrac{4(c_{0}-1)^{2}}{r^4}+\dfrac{8c_{2}(c_{0}-1)}{r^2}-\dfrac{10M}{9c_{2}c_{0}\mathcal{K}r}+\mathcal{O}(r^{0}).
\end{equation}
The curvature can be made small by choosing $c_{0}$ near 1, but as can be seen, it can not be made regular without sending $M$ to zero and recovering flat spacetime.\\
Starting from the near origin metric \eqref{originmetric} and asymptotic solution \eqref{eqfasympt}, we obtained the singular solutions with positive mass for the field equation \eqref{eqfiled2} and shown them in figure \eqref{ffrrplot00}b. As can be seen from the figure, unlike cubic gravity the plots are ascending and do not have a minimum due to the divergence of the coefficients of an arbitrary expansion of a metric function.

\begin{figure}[H]\hspace{0.4cm}
\centering
\subfigure{\includegraphics[width=0.5\columnwidth]{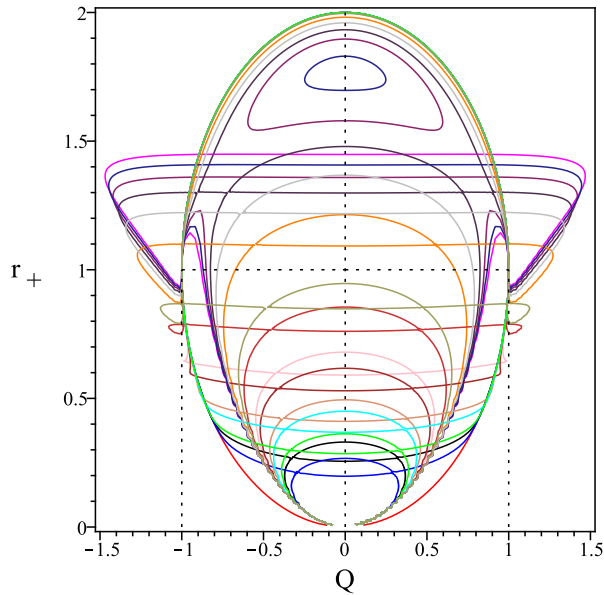}}
\caption{The plots of horizon radius $r_{+}$ in terms of $Q$ for large $-\mathcal{K}=\textcolor{red}{0}, 0.1, 0.5, 1, 10, 20, \textcolor{orange}{100}$.}
\label{ffrrplot123}
\end{figure}

Finally, we want to look at the equations \eqref{eqqfi1} and \eqref{eqqfi2} in the case of negative $\mathcal{K}$ and positive temperature $T>0$. For this case, as can be seen, from figure \eqref{ffrrplot123}, for every value of $Q$ there can be up to three solutions for $r_{+}$. But, these horizons cannot incorporate an asymptotically flat spacetime, because from the asymptotic metric function \eqref{eqfasympt}, in cases where $\mathcal{K}<0$, the homogeneous solution contains oscillating terms that spoil the asymptotic flatness.

\section{Conclusion}\label{con}
In this work, we studied the charged black hole solutions of Einsteinian quartic gravity --a modification to Einstein gravity that includes fourth-order curvature corrections-- in the presence of Maxwell electrodynamics. We find that similar to Einsteinian cubic gravity \cite{Frassino:2020zuv} the theory admits the charged black hole solutions with a charge greater than mass. 
We also find that, in the case of a positive coupling constant, there can be two asymptotically flat charged black holes with the same charge and mass. We show that only the larger black hole is thermodynamically stable because the smaller one has negative specific heat and free energy. The surprising feature of these charged black holes is the absence of an inner horizon, in contrast with the usual Reissner-Nordstrom solution. For the same values of mass and charge, there are two black hole solutions with different horizons.
Besides the black holes, we showed that there exists a naked singularity solution in pure (uncharged) Einsteinian quartic gravity.

\section*{Acknowledgements}
We would like to thank Ali Hajilou for useful discussions on Mathematica programming. 
SNS and SHH also thank the support of Iran National Science Foundation 99022223.

\appendix

\section{Explicit Constants}\label{appp1}
Here, we present the explicit constants that appear in \eqref{eqq155}:
\begin{align}
\mathcal{A}&={\it RootOf}( 20n{\it \_Z}^{5}\Lambda+( 30nf_{1}-15f_{1}) {\it \_Z}^{4}-30{\it \_Z}^{3}+(-60nM+90M) {\it \_Z}^{2}+\nonumber\\
&(12\mathcal{K}f_{1}^{4}-6n\mathcal{K}f_{1
}^{4}) {\it \_Z}+24\mathcal{K}f_{1}^{3}),  \nonumber
\\
\mathcal{B}&=80((( -\frac{3}{4}f_{{1}}^{2}+\Lambda) n+\frac{3}{8}
f_{1}^{2}) \mathcal{A}^{4}+(\frac{3}{4}f_{1}-3\Lambda nM
) \mathcal{A}^{3}-\frac{2}{5}f_{1}((\Lambda \mathcal{K}f_{1}^{3}-{
\frac {15}{4}}M) n+{\frac {45}{8}}M) \mathcal{A}^{2}\nonumber\\
&-\frac{4}{5}
(( -\frac{3}{16}f_{1}^{2}+\Lambda) n+\frac{3}{8}f_{1}^
{2}) f_{1}^{3}\mathcal{K}\mathcal{A}-\frac{3}{5}\mathcal{K}f_{1}^{4}) ( -\frac{3}{2}+n
) {n}^{-2}{\Lambda}^{-1} ( 2n-1) ^{-1},\nonumber\\
\mathcal{J}&=(-{\frac {3645}{1024}}\mathcal{K}-{\frac {135}{64}}\mathcal{K}^{2}{\Lambda}^{3}) {f_{{1}}}^{11}+{\frac {81}{64}}\mathcal{A}\Lambda( \mathcal{K}{
\Lambda}^{3}+{\frac {15}{8}}) \mathcal{K}{f_{{1}}}^{10}-\frac{3}{4}\Lambda
(( \mathcal{K}{\Lambda}^{4}+{\frac {135}{64}}\Lambda) \mathcal{A}^
{2}-{\frac {27}{4}}\mathcal{K}{\Lambda}^{3}-{\frac {405}{32}}) \mathcal{K}{f_{{1}
}}^{9}\nonumber\\
&+(({\frac {1485}{128}}\mathcal{K}{\Lambda}^{3}+{\frac {
18225}{1024}}+{\Lambda}^{6}\mathcal{K}^{2}) \mathcal{A}^{3}+(-3\mathcal{K}^{2}
{\Lambda}^{5}-{\frac {405}{64}}\mathcal{K}{\Lambda}^{2})\mathcal{A}+\frac{9}{2}{
\Lambda}^{2}M({\frac {405}{64}}+\mathcal{K}{\Lambda}^{3}) \mathcal{K}
) {f_{{1}}}^{8}+(({\frac {135}{32}}\mathcal{K}{\Lambda}^{
3}\nonumber\\
&+4{\Lambda}^{6}\mathcal{K}^{2}) \mathcal{A}^{2}-{\frac {1215}{32}}M
( \mathcal{K}{\Lambda}^{3}+{\frac {15}{16}}) \mathcal{A}-3\mathcal{K}^{2}{\Lambda}^
{5}-{\frac {405}{64}}\mathcal{K}{\Lambda}^{2}) {f_{{1}}}^{7}-{\frac {
225}{8}}\Lambda(({\frac {27}{16}}+\mathcal{K}{\Lambda}^{3}
) \mathcal{A}^{3}+(-\frac{4}{5}\mathcal{K}{\Lambda}^{3}M\nonumber\\
&-{\frac {27}{32}}M
 ) \mathcal{A}^{2}+(-{\frac {32}{225}}\mathcal{K}^{2}{\Lambda}^{5}-{
\frac {3}{20}}\mathcal{K}{\Lambda}^{2}) \mathcal{A}+3/2\mathcal{K}{\Lambda}^{2}M) 
{f_{{1}}}^{6}-{\frac {285}{8}}\Lambda(( \mathcal{K}{\Lambda}^{3
}+{\frac {675}{152}}) \Lambda M\mathcal{A}^{3}+{\frac {3}{38}}\mathcal{A}^{2
}{\Lambda}^{3}\mathcal{K}\nonumber\\
&+(-{\frac {405}{152}}M-{\frac {39}{19}}\mathcal{K}{
\Lambda}^{3}M) \mathcal{A}+{\frac {45}{38}}{\Lambda}^{3}\mathcal{K}{M}^{2}
 ) {f_{{1}}}^{5}+{\frac {135}{8}}(( \mathcal{K}{\Lambda}^{3
}+{\frac {15}{8}}) \mathcal{A}^{3}+(-{\frac {15}{4}}M-{\frac {
22}{9}}\mathcal{K}{\Lambda}^{3}M) \mathcal{A}^{2}+\nonumber\\
&\frac{11}{3}{M}^{2} ( {\frac {
405}{88}}+\mathcal{K}{\Lambda}^{3}) \mathcal{A}+\frac{1}{3}\mathcal{K}{\Lambda}^{2}M) {
\Lambda}^{2}{f_{{1}}}^{4}+30\mathcal{A}{\Lambda}^{2}M((\mathcal{K}{\Lambda
}^{4}+{\frac {135}{16}}\Lambda) \mathcal{A}^{2}-{\frac {45}{8}}\mathcal{A}M
\Lambda-{\frac {135}{64}}-\mathcal{K}{\Lambda}^{3}) {f_{{1}}}^{3}\nonumber\\
&+{\frac 
{2475}{8}}\mathcal{A}{\Lambda}^{3}M( \mathcal{A}^{2}\Lambda M-{\frac {15}{11}}
M+{\frac {3}{22}}\mathcal{A}) {f_{{1}}}^{2}-{\frac {225}{4}}\mathcal{A}{
\Lambda}^{4}M(\frac{15}{2}{M}^{2}+\mathcal{A}^{2}-4\mathcal{A}M) f_{{1}}-{
\frac {225}{2}}\mathcal{A}{\Lambda}^{4}{M}^{2}( \mathcal{A}^{2}\Lambda\nonumber\\
&- \frac{3}{2}\mathcal{A}M
\Lambda-\dfrac{1}{2}),\nonumber\\
\mathcal{G}&={\frac {891}{128}}( {\frac {405}{176}}+\mathcal{K}{\Lambda}^{3}) 
\mathcal{A}\mathcal{K}{f_{{1}}}^{11}-{\frac {27}{8}}((\mathcal{K}{\Lambda}^{4}+{
\frac {45}{16}}\Lambda) \mathcal{A}^{2}-\dfrac{5}{2}\mathcal{K}{\Lambda}^{3}-{\frac {
135}{32}}) \mathcal{K}{f_{{1}}}^{10}+\dfrac{3}{2}\mathcal{A}\Lambda((\mathcal{K}{
\Lambda}^{4}+{\frac {945}{256}}\Lambda) \mathcal{A}^{2}\nonumber\\
&-{\frac {27}{2
}}\mathcal{K}{\Lambda}^{3}-{\frac {1215}{32}}) \mathcal{K}{f_{{1}}}^{9}+( 
 (-{\frac {2835}{128}}\mathcal{K}{\Lambda}^{3}-{\frac {54675}{1024}}
 ) \mathcal{A}^{4}+({\frac {8505}{256}}\mathcal{K}{\Lambda}^{2}+9\mathcal{K}^{2
}{\Lambda}^{5}) \mathcal{A}^{2}-9{\Lambda}^{2}( \mathcal{K}{\Lambda}^{3}+
{\frac {3105}{256}}) M\mathcal{KA}\nonumber\\
&-{\frac {81}{4}}\,{K}^{2}{\Lambda}^{4}-
{\frac {1215}{32}}\mathcal{K}\Lambda) {f_{{1}}}^{8}+(( -{
\frac {18225}{1024}}-{\frac {945}{32}}\mathcal{K}{\Lambda}^{3}) \mathcal{A}^{3}
+{\frac {3645}{32}}M(\mathcal{K}{\Lambda}^{3}+\frac{5}{4}) \mathcal{A}^{2}+
 (18\mathcal{K}^{2}{\Lambda}^{5}+{\frac {8505}{128}}\mathcal{K}{\Lambda}^{2}
 ) \mathcal{A}\nonumber\\
 &-18{\Lambda}^{2}M({\frac {405}{64}}+\mathcal{K}{\Lambda}^{3}
 ) \mathcal{K}) {f_{{1}}}^{7}+{\frac {1755}{32}}\Lambda(( \mathcal{K}{\Lambda}^{3}+{\frac {675}{208}}) \mathcal{A}^{4}+(-{
\frac {38}{39}}\mathcal{K}{\Lambda}^{3}M-{\frac {315}{208}}M) \mathcal{A}^{3}
-{\frac {9}{13}}\mathcal{A}^{2}{\Lambda}^{2}\mathcal{K}\nonumber\\
&+{\frac {57}{13}}\mathcal{AK}{\Lambda}^
{2}M+{\frac {128}{585}}\mathcal{K}^{2}{\Lambda}^{4}+{\frac {6}{13}}\mathcal{K}
\Lambda) {f_{{1}}}^{6}+{\frac {135}{4}}\Lambda(
\Lambda M(K{\Lambda}^{3}+{\frac {45}{4}}) \mathcal{A}^{4}+
 ({\frac {45}{32}}+{\frac {11}{8}}\mathcal{K}{\Lambda}^{3}) \mathcal{A}^{
3}+(-{\frac {405}{32}}M\nonumber\\
&-7\mathcal{K}{\Lambda}^{3}M) \mathcal{A}^{2}+{
\frac {29}{8}}{\Lambda}^{2}\mathcal{K}(-{\frac {6}{29}}+{M}^{2}\Lambda
) \mathcal{A}+5\mathcal{K}{\Lambda}^{2}M) {f_{{1}}}^{5}-{\frac {135}{4}}
{\Lambda}^{2} (( \mathcal{K}{\Lambda}^{3}+{\frac {45}{8}}) \mathcal{A
}^{4}+(-{\frac {45}{4}}M\nonumber\\
&-\dfrac{10}{3}\mathcal{K}{\Lambda}^{3}M) \mathcal{A}^{3
}+( -{\frac {5}{12}}\mathcal{K}{\Lambda}^{2}+3{\Lambda}^{3}\mathcal{K}{M}^{2}+{
\frac {225}{8}}{M}^{2}) \mathcal{A}^{2}+{\frac {15}{4}}\mathcal{AK}{\Lambda}^
{2}M-5{\Lambda}^{2}\mathcal{K}{M}^{2}) {f_{{1}}}^{4}-{\frac {6075}{8}}
(\mathcal{A}^{4}M\Lambda\nonumber\\
&+({\frac {4}{135}}\mathcal{K}{\Lambda}^{3}-{
\frac {11}{18}}{M}^{2}\Lambda +\dfrac{1}{24}) \mathcal{A}^{3}+(-{\frac {
16}{135}}\mathcal{K}{\Lambda}^{3}M-\dfrac{1}{2}M) \mathcal{A}^{2}+{\frac {4}{45}}{
\Lambda}^{3}\mathcal{A}{M}^{2}\mathcal{K}+{\frac {4}{135}}\mathcal{K}{\Lambda}^{2}M) {
\Lambda}^{2}{f_{{1}}}^{3}\nonumber\\
&-{\frac {2025}{4}}\mathcal{A}^{2}{\Lambda}^{3}
(( -\dfrac{1}{8}+{M}^{2}\Lambda) \mathcal{A}^{2}+{\frac {19}{24}}
\mathcal{A}M-3{M}^{2}) {f_{{1}}}^{2}+{\frac {675}{2}}\mathcal{A}^{2}{\Lambda}
^{3}M(-\frac{1}{4}-{\frac {19}{8}}\mathcal{A}M\Lambda+3{M}^{2}\Lambda+\mathcal{A}^{2}
\Lambda) f_{{1}}\nonumber\\
&+{\frac {2025}{8}}\mathcal{A}^{2}{\Lambda}^{4}
( \mathcal{A}^{2}\Lambda M+(\frac{2}{9}-{M}^{2}\Lambda) \mathcal{A}-\frac{4}{3}M
 ),
 \end{align}

\begin{align}
\mathcal{H}&= -{\frac {405}{64}}\mathcal{AK} ( {\frac {135}{32}}+\mathcal{K}{\Lambda}^{3}
 ) {f_{{1}}}^{11}+{\frac {27}{16}}\mathcal{K}(( \mathcal{K}{\Lambda}^
{4}+{\frac {495}{64}}\Lambda) \mathcal{A}^{2}-13/2\mathcal{K}{\Lambda}^{3}-{
\frac {675}{32}}) {f_{{1}}}^{10}-{\frac {6075}{1024}}\mathcal{A}\Lambda
\mathcal{K}( -{\frac {512}{225}}\mathcal{K}{\Lambda}^{3}\nonumber\\
&+\mathcal{A}^{2}\Lambda-{\frac {
88}{5}}) {f_{{1}}}^{9}+ (( {\frac {273375}{4096}}+{
\frac {6885}{512}}\mathcal{K}{\Lambda}^{3}) \mathcal{A}^{4}-{\frac {6075}{128}}
\mathcal{A}^{2}{\Lambda}^{2}\mathcal{K}+{\frac {69255}{512}}\mathcal{AK}{\Lambda}^{2}M+{\frac 
{81}{4}}\mathcal{K}^{2}{\Lambda}^{4}+{\frac {3645}{32}}\mathcal{K}\Lambda) {f
_{{1}}}^{8}\nonumber\\
&+(( {\frac {4185}{128}}\mathcal{K}{\Lambda}^{3}+{
\frac {91125}{2048}}) \mathcal{A}^{3}-{\frac {12555}{128}}( {
\frac {1125}{496}}+\mathcal{K}{\Lambda}^{3}) M\mathcal{A}^{2}-{\frac {18225}{128}
}\mathcal{AK}{\Lambda}^{2}+{\frac {13365}{64}}\mathcal{K}{\Lambda}^{2}M) {f_{{1
}}}^{7}-{\frac {45}{2}}(( \mathcal{K}{\Lambda}^{3}\nonumber\\
&+{\frac {675}{
64}}) \mathcal{A}^{4}-{\frac {9}{8}}M({\frac {135}{32}}+\mathcal{K}{
\Lambda}^{3}) \mathcal{A}^{3}-{\frac {81}{32}}\mathcal{A}^{2}{\Lambda}^{2}\mathcal{K}+{
\frac {213}{16}}\mathcal{AK}{\Lambda}^{2}M+{\frac {81}{16}}\mathcal{K}\Lambda) 
\Lambda\,{f_{{1}}}^{6}-{\frac {42525}{128}}\Lambda( \mathcal{A}^{4}M
\Lambda\nonumber\\
&+( {\frac {8}{63}}\mathcal{K}{\Lambda}^{3}+3/7) \mathcal{A}^{3}+
 ( -{\frac {16}{35}}\mathcal{K}{\Lambda}^{3}M-{\frac {15}{7}}M) 
\mathcal{A}^{2}+{\frac {8}{35}}\mathcal{K}{\Lambda}^{2}( -1+{M}^{2}\Lambda
) \mathcal{A}+{\frac {16}{15}}\mathcal{K}{\Lambda}^{2}M) {f_{{1}}}^{5}\nonumber\\
&+{
\frac {18225}{64}}{\Lambda}^{2}( \mathcal{A}^{4}-2\mathcal{A}^{3}M+( -
{\frac {32}{405}}\mathcal{K}{\Lambda}^{2}+{\frac {23}{6}}{M}^{2}) \mathcal{A}
^{2}+{\frac {8}{15}}\mathcal{AK}{\Lambda}^{2}M-{\frac {8}{15}}\Lambda
({M}^{2}\Lambda-\frac{2}{9}) \mathcal{K}) {f_{{1}}}^{4}+\nonumber\\
&{\frac {
10125}{16}}( \mathcal{A}^{4}M\Lambda+( -\frac{3}{5}{M}^{2}\Lambda+{
\frac {9}{40}}) \mathcal{A}^{3}-{\frac {9}{8}}M\mathcal{A}^{2}-{\frac {4}{225
}}\mathcal{AK}{\Lambda}^{2}+{\frac {4}{25}}\mathcal{K}{\Lambda}^{2}M) {\Lambda}
^{2}{f_{{1}}}^{3}+{\frac {6075}{32}}(( {M}^{2}\Lambda-2
/3) {A}^{2}\nonumber\\
&+11/3\mathcal{A}M-9{M}^{2}) \mathcal{A}^{2}{\Lambda}^{3}{f_
{{1}}}^{2}-{\frac {2025}{8}}\mathcal{A}^{2}{\Lambda}^{3}( \mathcal{A}^{2}
\Lambda M+( -9/4{M}^{2}\Lambda+1/6 ) \mathcal{A}-5/6 M+9/4{M}^
{3}\Lambda) f_{{1}}\nonumber\\
&+{\frac {225}{16}}{\Lambda}^{4}{\mathcal{A}}^{2}
( \mathcal{A}-3M)( -9M+\mathcal{A}),\nonumber\\
 \mathcal{N}&=-{\frac {16}{25}}\mathcal{A}(\mathcal{K}{\Lambda}^{3}+{\frac {3375}{256}}
) \mathcal{K}{f_{{1}}}^{10}+(-{\frac {32}{25}}\mathcal{K}^{2}{\Lambda}^{
3}+{\frac {63}{20}}\Lambda\mathcal{A}^{2}\mathcal{K}-{\frac {27}{2}}\mathcal{K}) {f_{
{1}}}^{9}-{\frac {39}{40}}\mathcal{A}\Lambda\mathcal{K}(-{\frac {420}{13}}+\mathcal{A}^
{2}\Lambda) f_{1}^{8}\nonumber\\
&+(( \mathcal{K}{\Lambda}^{3}+{
\frac {135}{8}}) \mathcal{A}^{4}-{\frac {39}{4}}\mathcal{A}^{2}{\Lambda}^{2}\mathcal{K}
+{\frac {513}{20}}\mathcal{AK}{\Lambda}^{2}M+{\frac {216}{5}}\mathcal{K}\Lambda
 ) f_{1}^{7}+(( {\frac {18}{5}}\mathcal{K}{\Lambda}^{3}
+{\frac {135}{8}}) \mathcal{A}^{3}+ (-{\frac {48}{5}}\mathcal{K}{\Lambda}
^{3}M\nonumber\\
&-{\frac {135}{2}}M) \mathcal{A}^{2}-39\mathcal{AK}{\Lambda}^{2}+{\frac {
234}{5}}\mathcal{K}{\Lambda}^{2}M ) {f_{{1}}}^{6}+( {\frac {99}{4}
}\Lambda\mathcal{A}^{3}M-{\frac {192}{5}}\mathcal{AK}{\Lambda}^{3}M+8\mathcal{A}^{2}{
\Lambda}^{3}\mathcal{K}-{\frac {225}{4}}\mathcal{A}^{4}\Lambda-\nonumber\\
&{\frac {216}{5}}\mathcal{K}{
\Lambda}^{2}) {f_{{1}}}^{5}-48\Lambda( 6/5\mathcal{K}{\Lambda
}^{2}M-1/3\mathcal{AK}{\Lambda}^{2}+\mathcal{A}^{4}M\Lambda-{\frac {69}{16}}M\mathcal{A}^{2}
+{\frac {9}{8}}\mathcal{A}^{3}) {f_{{1}}}^{4}+60{\Lambda}^{2}
 ({\frac {16}{75}}\mathcal{K}\Lambda+\mathcal{A}^{4}\nonumber\\
 &-2\mathcal{A}^{3}M+{\frac {33}{10}
}\mathcal{A}^{2}{M}^{2}) {f_{{1}}}^{3}+( 60\mathcal{A}^{4}{\Lambda}^{
3}M+( -36{\Lambda}^{3}{M}^{2}+54{\Lambda}^{2}) \mathcal{A}^{3
}-192\mathcal{A}^{2}M{\Lambda}^{2}) {f_{{1}}}^{2}-20\mathcal{A}^{2}{\Lambda
}^{3}( -{\frac {27}{5}}\mathcal{A}M\nonumber\\
&+\mathcal{A}^{2}+{\frac {54}{5}}{M}^{2}
 ) f_{{1}}-16\mathcal{A}^{2}{\Lambda}^{3}( \mathcal{A}-3M),\nonumber\\
 \mathcal{P}&={\frac {9}{16}}{f_{{1}}}^{8}\mathcal{AK}+(\mathcal{K}-{\frac {17}{120}}\Lambda
\mathcal{A}^{2}\mathcal{K}) {f_{{1}}}^{7}+1/45\mathcal{A}\Lambda\mathcal{K}( -{\frac {153
}{2}}+\mathcal{A}^{2}\Lambda) {f_{{1}}}^{6}+({\frac {4}{15}}\mathcal{A
}^{2}{\Lambda}^{2}\mathcal{K}-2/3\mathcal{AK}{\Lambda}^{2}M-8/3\mathcal{K}\Lambda\nonumber\\
&-{\frac {15}{16
}}\mathcal{A}^{4}) {f_{{1}}}^{5}+( -5/4\mathcal{A}^{3}+{\frac {35}{8}
}M\mathcal{A}^{2}+4/3\mathcal{AK}{\Lambda}^{2}-4/3\mathcal{K}{\Lambda}^{2}M) {f_{{1}}
}^{4}+5/2\Lambda( {\frac {32}{45}}\mathcal{K}\Lambda-{\frac {13}{30}
}\mathcal{A}^{3}M+\mathcal{A}^{4}) {f_{{1}}}^{3}\nonumber\\
&+( \mathcal{A}^{2}\Lambda M-11
M+10/3\mathcal{A}) \mathcal{A}^{2}\Lambda{f_{{1}}}^{2}-5/3\mathcal{A}^{2}{\Lambda
}^{2}( \mathcal{A}^{2}+3{M}^{2}-2\mathcal{A}M) f_{{1}}-{\frac {20}{9}}
\mathcal{A}^{2}{\Lambda}^{2}( \mathcal{A}-3M), \nonumber\\
\mathcal{S}&={\frac {9}{20}}{f_{{1}}}^{6}\mathcal{AK}-{\frac {2}{35}}\mathcal{K}( \mathcal{A}^{2}
\Lambda-15) {f_{{1}}}^{5}-4/5\mathcal{AK}\Lambda{f_{{1}}}^{4}+
(-{\frac {9}{14}}\mathcal{A}^{4}-{\frac {48}{35}}\mathcal{K}\Lambda) {
f_{{1}}}^{3}-{\frac {15}{14}}\mathcal{A}^{2}(-{\frac {16}{5}}M+\mathcal{A}
) {f_{{1}}}^{2}\nonumber\\
&+\mathcal{A}^{3}\Lambda( \mathcal{A}-3/7M) f_{{1}
}+{\frac {12}{7}}\mathcal{A}^{2}\Lambda( \mathcal{A}-3M),\nonumber\\
\mathcal{T}&=-{\frac {27}{8}}{f_{{1}}}^{7}\mathcal{K}+9/4{f_{{1}}}^{6}\mathcal{AK}\Lambda-3/2{f_{
{1}}}^{5}\mathcal{A}^{2}{\Lambda}^{2}\mathcal{K}+(( \mathcal{K}{\Lambda}^{3}+{\frac {
135}{8}}) \mathcal{A}^{3}+3\mathcal{K}{\Lambda}^{2}M) {f_{{1}}}^{4}-{
\frac {135}{4}}\mathcal{A}M{f_{{1}}}^{3}\nonumber\\
&+{\frac {45}{2}}{f_{{1}}}^{2}\mathcal{A}^{2}
\Lambda M-30f_{{1}}{\Lambda}^{2}\mathcal{A}^{3}M+30\mathcal{A}{M}^{2}{\Lambda}^{2},\nonumber\\
\mathcal{U}&={\frac {189}{16}}\mathcal{AK}{f_{{1}}}^{7}-{\frac {27}{4}}\mathcal{K}( -2+{A}^{2
}\Lambda) {f_{{1}}}^{6}+{\frac {15}{4}}\mathcal{A}\Lambda( \mathcal{A}^
{2}\Lambda-{\frac {18}{5}}) \mathcal{K}{f_{{1}}}^{5}+(( -{
\frac {135}{4}}-2\mathcal{K}{\Lambda}^{3}) \mathcal{A}^{4}+15/2\mathcal{A}^{2}{
\Lambda}^{2}\mathcal{K}\nonumber\\
&-21/2\mathcal{AK}{\Lambda}^{2}M) {f_{{1}}}^{4}+( 
 (-{\frac {135}{8}}-4\mathcal{K}{\Lambda}^{3}) \mathcal{A}^{3}+{\frac {
405}{4}}M\mathcal{A}^{2}-12\mathcal{K}{\Lambda}^{2}M) {f_{{1}}}^{3}+{\frac {
135}{4}}\mathcal{A}^{3}\Lambda( -5/3M+\mathcal{A}) {f_{{1}}}^{2}\nonumber\\
&+
 ( 60\mathcal{A}^{4}{\Lambda}^{2}M-45\mathcal{A}^{2}\Lambda\,M) f_{{1}
}+30\mathcal{A}^{2}{\Lambda}^{2}M( \mathcal{A}-3\,M),
\end{align}
\begin{align}
\mathcal{V}&=-{\frac {405}{32}}\mathcal{AK}{f_{{1}}}^{7}+{\frac {81}{16}}\mathcal{K} ( \mathcal{A}^{2}
\Lambda-4) {f_{{1}}}^{6}-3/2\mathcal{A}\Lambda( \mathcal{A}^{2}\Lambda-
{\frac {27}{2}}) \mathcal{K}{f_{{1}}}^{5}+( 18\mathcal{K}\Lambda+9\mathcal{AK}{
\Lambda}^{2}M+{\frac {405}{16}}\mathcal{A}^{4}\nonumber\\
&-6\mathcal{A}^{2}{\Lambda}^{2}\mathcal{K}
 ) {f_{{1}}}^{4}+( -{\frac {405}{4}}\,M\mathcal{A}^{2}+{\frac {405
}{16}}\mathcal{A}^{3}+18\mathcal{K}{\Lambda}^{2}M-6\mathcal{AK}{\Lambda}^{2}) {f_{{1}
}}^{3}-{\frac {135}{4}}\mathcal{A}^{3}\Lambda( \mathcal{A}-7/6\,M) {f_{
{1}}}^{2}\nonumber\\
&-{\frac {45}{2}}\mathcal{A}^{2}\Lambda( \mathcal{A}^{2}\Lambda\,M-4
\,M+\mathcal{A}) f_{{1}}+15/2\mathcal{A}^{2}{\Lambda}^{2}( \mathcal{A}-3\,M
 ) ^{2},\nonumber\\
 \mathcal{W}&={\frac {13}{20}}{f_{{1}}}^{6}\mathcal{AK}-2/15\mathcal{K}( -9+\mathcal{A}^{2}\Lambda
 ) {f_{{1}}}^{5}-4/5\mathcal{AK}\Lambda\,{f_{{1}}}^{4}+ ( -{\frac {
16}{15}}\mathcal{K}\Lambda-\mathcal{A}^{4}) {f_{{1}}}^{3}+( 5\,M\mathcal{A}^{2}-3
/2\mathcal{A}^{3}) {f_{{1}}}^{2}\nonumber\\
&+\mathcal{A}^{3}\Lambda( \mathcal{A}-M) 
f_{{1}}+4/3\mathcal{A}^{2}\Lambda( \mathcal{A}-3\,M),\nonumber\\
\mathcal{X}&=-4/5\,{f_{{1}}}^{4}\mathcal{AK}-8/5\mathcal{K}{f_{{1}}}^{3}+f_{{1}}\mathcal{A}^{4}+2\mathcal{A}^{2}
 \left( \mathcal{A}-3\,M \right).
\end{align}
\section{Near Horizon Solutions}\label{appp2}
Here, we present additional terms that appear in the near horizon solution \eqref{eq7}:
\begin{align}
f_{3}&=-\dfrac{1}{72f_{
1}^{2}\mathcal{K}r_{+}^{2}(2+r_{+}f_{1})}(-20r_{+}^{5}f_{2}-20\,\Lambda\,r_{+}^{5}-20\,r_{+}^{4}f_{1}+96\,\mathcal{K}f_{1}^{2}r_{+}^{3
}f_{2}^{2}+96\,f_{1}^{3}\mathcal{K}+\nonumber\\
&96\,\mathcal{K}f_{1}r_{+}^{2}f_{2}^
{2}-152\,\mathcal{K}f_{1}^{3}r_{+}^{2}f_{2}+20\,r_{+}\,Q^{2}+58
\,r_{+}f_{1}^{4}\mathcal{K}-240\mathcal{K}f_{1}^{2}r_{+}f_{2}),\\
f_{4}&=\dfrac{1}{3456r_{+}
^{3}f_{1}^{4}\mathcal{K}^{2}(2+r_{+}\,f_{1}) ^{2}}(-6912\,f_{1}^{5}\mathcal{K}^{2}-9744\,f_{1}
^{6}\mathcal{K}^{2}r_{+}+1910{r_{+}}^{5}f_{1}^{4}\mathcal{K}-3464\,{r_{+}
}^{2}f_{1}^{7}\mathcal{K}^{2}\nonumber\\
&+100\,r_{+}^{9}f_{2}+23040\,\mathcal{K}^{2}r_{+}^{4}f_{2}^{3}f_{1}^{3}+11520\,\mathcal{K}^{2}r_{+}^{5}f_
{2}^{3}f_{1}^{4}-3840\,r_{+}^{6}f_{2}^{2}f_{1}\mathcal{K}+
13824\,\mathcal{K}^{2}r_{+}^{3}f_{2}^{3}f_{1}^{2}+\nonumber\\
&1680\,r_{+}^{5}f_{2}f_{1}^{2}\mathcal{K}-3360\,r_{+}^{6}f_{2}f_{1}\mathcal{K}\Lambda+
27648\,r_{+}\,f_{2}f_{1}^{4}\mathcal{K}^{2}-55680\,r_{+}^{3}f_{
2}^{2}f_{1}^{4}\mathcal{K}^{2}+40560\,r_{+}^{2}f_{2}f_{1}^{5
}\mathcal{K}^{2}\nonumber\\
&-34560\,r_{+}^{2}f_{2}^{2}f_{1}^{3}\mathcal{K}^{2}-3600\,\mathcal{
K}f_{1}^{2}r_{+}^{7}f_{2}^{2}+200\,\mathcal{K}f_{1}^{3}r_{+}^{6}f_{2}-23040\,\mathcal{K}^{2}f_{1}^{5}r_{+}^{4}f_{2}^{2}+
15112\,\mathcal{K}^{2}f_{1}^{6}r_{+}^{3}f_{2}\nonumber\\
&+2320\,r_{+}^{6}
f_{1}^{3}\mathcal{K}\Lambda+3360\,f_{1}^{2}\mathcal{K}\Lambda\,r_{+}^{5}-3120
\,\mathcal{K}f_{1}^{2}r_{+}^{7}f_{2}\Lambda+100\,r_{+}^{9}
\Lambda+100\,r_{+}^{8}f_{1}-1840\,r_{+}^{2}{Q}^{2}f_{1}
^{3}\mathcal{K}\nonumber\\
&-2400\,r_{+}\,{Q}^{2}f_{1}^{2}\mathcal{K}+3360\,r_{+}^{2}f_{2
}f_{1}\mathcal{K}{Q}^{2}+3120\,\mathcal{K}f_{1}^{2}r_{+}^{3}f_{2}{Q}^{2}-100
\,r_{+}^{5}{Q}^{2}+2640\,f_{1}^{3}\mathcal{K}r_{+}^{4}).
\end{align}

\end{document}